\documentclass[11pt,twoside]{article}
\usepackage{amsmath,amssymb,theorem}
\def\notepagedim{\textwidth 6in \textheight 8in \topmargin 0in
\oddsidemargin 0.25in \evensidemargin  \oddsidemargin}
\notepagedim

\newcommand{\donothing}[1]{{}}

\numberwithin{equation}{section}
\newtheorem{theorem}{Theorem}[section]

\newtheorem{lemma}[theorem]{Lemma}

\newtheorem{definition}[theorem]{Definition}

\newtheorem{remark}[theorem]{Remark}
\newenvironment{proof}{{\bf Proof}.\ }{ \hfill $\square$}
\newcommand{\mat}[1]{\begin{bmatrix} #1 \end{bmatrix}}

\newcommand{\R}{\mathbb{R}}
\newcommand{\C}{\mathbb{C}}

\newcommand{\pd}{\partial}

\begin{document}

\title{Survey on global existence in the nonlinear Dirac equations \\ in one dimension}

\author{Dmitry Pelinovsky \\
{\small Department of Mathematics, McMaster
University, Hamilton, Ontario, Canada, L8S 4K1}
}

\date{\today}
\maketitle

\begin{abstract}
We consider the nonlinear Dirac equations in one dimension and
review various results on global existence of solutions in $H^1$.
Depending on the character of the nonlinear terms, existence of the large-norm solutions
can be extended for all times. Global existence of the small-norm solutions
is proved for the most general nonlinear Dirac equations with cubic and higher-order nonlinear terms. 
Integrability of the massive Thirring model is used to find conditions that no solitons 
occur in the Cauchy problem with small initial data in a subspace of $L^2$.
\end{abstract}

\section{Introduction}

The goal of this article is to survey a number of recent results on
global well-posedness of the nonlinear Dirac equations in the space of one
dimension. The nonlinear Dirac equations are known for long time in quantum mechanics and 
relativity theory \cite{gross-neveu,Thirring}.
Recently these equations were used to model other physical systems such as
photonic crystals and Bose--Einstein condensates in optical lattices
\cite{AP,Pelinovsky,SS}.

The nonlinear Dirac equations are similar to the nonlinear Klein-Gordon equation on one hand
and to the nonlinear Schr\"{o}dinger equation on the other hand. In the former case,
the reduction is possible for a special form of nonlinear terms as the linear dispersion relation
between the two models are identical. In the latter case, the reduction appears in the asymptotic
limit using an envelope wave approximation near one branch of the wave spectrum. Since
analysis of global well-posedness in the nonlinear Klein-Gordon and nonlinear Schr\"{o}dinger equations
has been booming in the last ten years, it is not surprising that the interest of harmonic and
PDE analysts turns recently to the nonlinear Dirac equations.

The organization of this article is as follows. Section 2 sets up the nonlinear Dirac equations and reviews
a number of physically relevant models. Global well-posedness in $H^1$ is studied in
Section 3 using apriori estimates as in Goodman {\em et al.} \cite{GWH}. We show that the nonlinear Dirac equations
with a special structure of nonlinear terms are globally well-posed even for large
powers of the nonlinear terms. This result is different from the behavior of the nonlinear Schr\"{o}dinger equation.

Section 4 deals with global solutions for small initial data in $H^1$ using analysis in Strichartz spaces
as in Pelinovsky \& Stefanov \cite{PS}.  We prove the global existence of
small-norm solutions that scatter to zero for the nonlinear Dirac
equations with quintic and higher-order nonlinear terms.

The nonlinear Dirac equations with cubic terms are considered in Section 5. Scattering to zero
for small-norm solutions is proved using the recent results of Hayashi \& Naumkin \cite{HN1,HN2}.
We also discuss the special role of the integrable version of the nonlinear Dirac equations
known as the massive Thirring model \cite{Thirring}, for which global well-posedness in $L^2$ was proved
recently by Candy \cite{Candy}. Using the formalism of the inverse scattering transform, 
we show that small initial data in a subspace of $L^2$ are associated to purely continuous spectrum of 
the Lax operator and admits no solitons in the long-time asymptotics. 

{\bf Acknowledgments}: Part of this work was done during the visits
of the author to University of British Columbia and the Kyoto University in 2010.
The author would like to thank K. Nakanishi and T.P. Tsai for useful discussions regarding
this article.

\section{Model}

Let us consider the nonlinear Dirac equations
\begin{equation}
\label{Dirac1}
\left\{ \begin{array}{cc} i (u_t + u_x) + v = \partial_{\bar{u}} W(u,v), \\
i (v_t - v_x) + u = \partial_{\bar{v}} W(u,v), \end{array} \right.
\end{equation}
where $(x,t) \in \mathbb{R}^2$, $(u,v) \in \mathbb{C}^2$, and $W(u,v) : \C^2 \to \R$ is a
nonlinear function which satisfies the following three conditions:
\begin{itemize}
\item symmetry $W(u,v) = W(v,u)$; \item gauge invariance $W(e^{i
\theta} u,e^{i \theta} v) = W(u,v)$ for any $\theta \in \R$; \item
polynomial in $(u,v)$ and $(\bar{u},\bar{v})$.
\end{itemize}

The nonlinear Dirac equations can be rewritten in the abstract
evolutionary form,
\begin{equation}
\label{Dirac2} i \pd_t {\bf u} = {\cal H} {\bf u} + {\bf
f}({\bf u}), \quad {\cal H}
= \mat{-i \pd_x & -1 \\ -1 & i \pd_x}, \quad {\bf u} = \left[ \begin{array}{c}
u \\ v \end{array} \right], \quad 
{\bf f}({\bf u}) = \left[ \begin{array}{c}
\partial_{\bar{u}} W(u,v) \\ \partial_{\bar{v}} W(u,v) \end{array} \right].
\end{equation}

A homogeneous quartic polynomial $W(u,v)$ satisfying the three properties above is characterized
by Chugunova \& Pelinovsky \cite{Chu-Pel},
\begin{eqnarray}
W = \alpha_1 (|u|^4 + |v|^4) + \alpha_2 |u|^2 |v|^2 +
\alpha_3 (\bar{u} v + u \bar{v})^2 + \alpha_4 (|u|^2 + |v|^2)
(\bar{u} v + u \bar{v}), \label{potential}
\end{eqnarray}
where $(\alpha_1,\alpha_2,\alpha_3,\alpha_4) \in \R^4$ are numerical coefficients.

The standard example of the nonlinear Dirac equations with
\begin{eqnarray}
\label{coupled-mode}
W = \alpha (|u|^2 + |v|^2)^2 + 2 \alpha |u|^2 |v|^2, \quad \alpha \in \R,
\end{eqnarray}
occurs in the context of periodic dielectric materials
under the Bragg resonance \cite{SS}.
The account of zero-mean periodic modulation of the nonlinear refractive index
gives the nonlinear Dirac equations with
\begin{eqnarray}
W = \alpha (\bar{u} v + u \bar{v}) (|u|^2 + |v|^2) + \beta ((\bar{u} v + u \bar{v})^2 - 2|u|^2 |v|^2),
\end{eqnarray}
where $\alpha \in \R$ and $\beta \in \R$ are proportional to different Fourier coefficients of
the nonlinear refractive index \cite{AP}.

Two models are relevant for general relativity: the massive Thirring model when $W = 4 |u|^2 |v|^2$ \cite{Thirring}
and the massive Gross--Neveu model when $W = 2 (\bar{u} v + u \bar{v})^2$ \cite{gross-neveu}.
Under the following change of variables,
\begin{equation}
\label{Dirac3}
\mbox{\boldmath $\psi$} = T {\bf u}, \quad {\bf g} = T {\bf f},\quad T =\mat{ 1 &
-1 \\ -i & -i},
\end{equation}
the nonlinear Dirac equation (\ref{Dirac2}) can be written in the equivalent form
\begin{equation}
\label{Dirac4}
i \pd_t \mbox{\boldmath $\psi$} = {\cal M} \mbox{\boldmath $\psi$} + {\bf g}, \quad {\cal M} = \mat{1 & \pd_x \\ -\pd_x & -1}.
\end{equation}
If $\mbox{\boldmath $\psi$} = (\psi,\phi)$, system (\ref{Dirac4}) can be written as follows: 
the massive Thirring model is 
\begin{equation}
\label{MTM}\left\{ \begin{array}{cc}
i \psi_t - \psi - \phi_x = (\psi^2 + \phi^2) \bar{\psi}, \\
i \phi_t + \phi + \psi_x = (\psi^2 + \phi^2) \bar{\phi}, \end{array} \right.
\end{equation}
and the massive Gross--Neveu model is
\begin{equation}
\label{MGN}\left\{ \begin{array}{cc}
i \psi_t - \psi - \phi_x = (\psi^2 - \phi^2) \psi, \\
i \phi_t + \phi + \psi_x = (\phi^2 - \psi^2) \phi, \end{array} \right.
\end{equation}

Thus, we see that all coefficients of the general quartic potential (\ref{potential})
have physically relevant applications. Note also that $W$ may also have sixth-order
and higher-order terms, as in the context of the Feshbach
resonance for Bose--Einstein condensates \cite{ChPorPel}, where
$$
W = \alpha (|u|^2 + |v|^2) |u|^2 |v|^2, \quad \alpha \in \R.
$$

Local existence of solutions of the nonlinear Dirac system in Sobolev space $H^s(\R)$ can be proved with
standard methods using the Duhamel formulation and the fixed-point arguments \cite{Del,GWH}.
If ${\bf u}_0 \in H^s(\R)$ for a fixed $s > \frac{1}{2}$, then there exists
a $T > 0$ such that the nonlinear Dirac equations
(\ref{Dirac2}) admits a unique solution
$$
{\bf u}(t) \in C([0,T],H^s(\R)) \cap C^1([0,T],H^{s-1}(\R)),
$$
where ${\bf u}(t)$ depends continuously on the initial data ${\bf
u}(0) = {\bf u}_0$.

In what follows, we review results on global well-posedness
of general nonlinear Dirac equations (\ref{Dirac1}) in some subspaces of $H^1(\R)$ 
or $L^2(\R)$.

\section{Global well-posedness in $H^1$}

There exist three conserved quantities of the nonlinear Dirac equations (\ref{Dirac1})
corresponding to {\em Hamiltonian} $H$, {\em momentum} $P$, and {\em charge} $Q$,
\begin{equation}
\label{quantity-H} H = \frac{i}{2} \int_{\mathbb{R}} \left(
u_x \bar{u} - u \bar{u}_x - v_x \bar{v} + v \bar{v}_x \right) dx +
\int_{\R} \left( v \bar{u} + u \bar{v} - W(u,v) \right) dx,
\end{equation}
\begin{equation}
\label{quantity-P} P = \frac{i}{2} \int_{\mathbb{R}} \left(
u \bar{u}_x - u_x \bar{u} + v \bar{v}_x - v_x \bar{v} \right) dx,
\end{equation}
and
\begin{equation}
\label{quantity-Q} Q = \int_{\mathbb{R}} \left( |u|^2+|v|^2
\right) dx.
\end{equation}
These conserved quantities are well defined for a local solution in $H^s(\R)$ for $s > \frac{1}{2}$
thanks to the Banach algebra of $H^s(\R)$, $s > \frac{1}{2}$ with respect to multiplication.

Unlike the nonlinear Schr\"{o}dinger equation, the
Hamiltonian $H$ is not useful for analysis of global well-posedness because
the quadratic part of $H$ is sign-indefinite. Nevertheless, for a special nonlinear
function $W(u,v)$, local solutions can be extended to global solutions for all $t \in \R$.
The following theorem generalizes the result of Delgado \cite{Del}
for the massive Thirring model (\ref{MTM}) and the result of
Goodman {\em et al.} \cite{GWH} for the nonlinear Dirac equations with
$W$ in (\ref{coupled-mode}).

\begin{theorem}
\label{theorem-global-wellposedness-cme} Assume that $W$ is a
polynomial in variables $|u|^2$ and $|v|^2$. Let ${\bf u}(t) \in C([0,T],H^1(\R))$ be
a local solution of the nonlinear Dirac equations (\ref{Dirac1}) for some $T > 0$. Then,
${\bf u}(t) \in C(\R,H^1(\R))$.
\end{theorem}

\begin{proof}
To extend the local solution ${\bf u}(t) \in C([0,T],H^1(\R))$ for all $t \in \R$, it is
sufficient to prove that the $H^1$-norm of the solution ${\bf u}(t)$ satisfies the estimate
\begin{equation}
\label{bound-T}
\sup_{t \in [0,T]} \| {\bf u}(t) \|_{H^1} \leq C(T),
\end{equation} where the constant $C(T)$ is finite for $T <
\infty$ but may grow as $T \to \infty$. By the
conservation of $Q$ in (\ref{quantity-Q}), we have
\begin{equation}
\| {\bf u}(t) \|_{L^2} = \| {\bf u}_0 \|_{L^2}, \quad t
\in \mathbb{R}.
\end{equation}

To consider $\| {\bf u}(t) \|_{L^{2p+2}}$ for a fixed $p > 0$, 
we multiply the first equation of system (\ref{Dirac1})
by $|u|^{2p} \bar{u}$ and the second equation by $|v|^{2p} \bar{v}$, add
the two equations, and take the imaginary part. If $W$ depends
only on $|u|^2$ and $|v|^2$, the nonlinear function is cancelled
out and we obtain
$$
\frac{1}{p+1} \partial_t \left( |u|^{2p+2} + |v|^{2p+2} \right) +
\frac{1}{p+1} \partial_x \left( |u|^{2p+2} - |v|^{2p+2} \right) =
{\rm i} (v \bar{u} - \bar{v} u) (|u|^{2p} - |v|^{2p}).
$$
Integrating this balance equation on $x\in \mathbb{R}$ for a local
solution in $C([0,T],H^1(\R))$ and using inequality $|u||v| \leq
\frac{1}{2} (|u|^2 + |v|^2)$, we obtain apriori estimate
\begin{equation}
\frac{d}{d t} \| {\bf u}(t) \|^{2p+2}_{L^{2p+2}} \leq 4 (p+1) \|
{\bf u}(t) \|^{2p+2}_{L^{2p+2}}.
\end{equation}
By Gronwall's inequality, we
have
\begin{equation}
\label{bound-L-p} \|{\bf u}(t) \|_{L^{2p+2}} \leq e^{2 |t|} \|
{\bf u}_0 \|_{L^{2p+2}}, \quad t \in [0,T].
\end{equation}
Since the estimate holds for any $p > 0$, it holds for $p \to
\infty$ and gives apriori estimate on the $L^{\infty}$-norm of
the local solution ${\bf u}(t)$. The bound on the $L^{\infty}$-norm is
needed to control the growth rate of the $L^2$-norm of ${\bf u}_x(t)$ as $t \to \infty$.

Taking $x$-derivatives and performing a similar computation,
we obtain the balance equation
$$
\partial_t \left( |u_x|^2 + |v_x|^2 \right) +
\partial_x \left( |u_x|^2 - |v_x|^2 \right) =
{\rm i} \left( u_x \partial_x \partial_{\bar{u}} + v_x \partial_x
\partial_{\bar{v}} - \bar{u}_x
\partial_x \partial_u - \bar{v}_x \partial_x
\partial_v \right) W(u,v).
$$
Let $N \geq 1$ be the degree of the polynomial $W$ in variables $|u|^2$ and $|v|^2$.
Integrating the previous equation over $x \in \R$ and using bound
(\ref{bound-L-p}), we obtain the estimate
$$
\frac{d}{d t} \| {\bf u}_x(t) \|^2_{L^2} \leq C_W e^{4(N-1)
|t|} \| {\bf u}_x(t) \|^2_{L^2},
$$
where the constant $C_W > 0$ depends on the coefficients of the polynomial $W$.
By Gronwall's inequality again, we obtain
\begin{equation}
\| {\bf u}_x(t) \|^2_{L^2} \leq  e^{\frac{C_W}{4(N-1)}
(e^{4(N-1)|t|} - 1)} \| ({\bf u}_0)_x \|^2_{L^2}, \quad t \in [0,T].
\end{equation}
The exponential factor remains bounded for any finite
time $T > 0$. Therefore, $C(T) < \infty$ if $T < \infty$ and bound (\ref{bound-T}) gives global well-posedness
of the nonlinear Dirac equations in the $H^1$-norm.
\end{proof}

\begin{remark}
The result of Theorem \ref{theorem-global-wellposedness-cme} is very different from
the behavior of the nonlinear Schr\"{o}dinger equations, where global solutions may not exist for
nonlinear terms generated by a polynomial $W$ in variables $|u|^2$ and $|v|^2$ of degree $N \geq 3$.
\end{remark}

If $W$ also depends on $(\bar{u}v + u \bar{v})$, apriori
estimates of the $L^p$-norm include nonlinear terms which may lead
to the finite-time blow-up of solutions in $L^{\infty}$ and $H^1$
norms. That is, there may exist $T_{\rm max} < \infty$ such that
\begin{equation}
\label{finite-time-blow-up}
\lim_{t \uparrow T_{\rm max}} \| {\bf u} \|_{H^1} = \infty.
\end{equation}
The next sections address the question if the finite-time blow-up
(\ref{finite-time-blow-up}) can be excluded at least for small-norm solutions
in the general nonlinear Dirac equations (\ref{Dirac1}).

\section{Global well-posedness in Strichartz spaces}

Strichartz spaces $L^p_t L^q_x$ and $L^q_x L^p_t$ are defined
for $1 \leq p,q \leq \infty$ by the norms
\begin{eqnarray}
\label{Strichartz}
\|f \|_{L^p_t L^q_x} := \left( \int_0^T \| f(\cdot,t) \|_{L^q_x}^p dt \right)^{1/p}, \quad
\|f \|_{L^q_x L^p_t} := \left( \int_{\R} \| f(x,\cdot) \|_{L^p_t}^q dx \right)^{1/q},
\end{eqnarray}
where $T > 0$ is an arbitrary time including $T = \infty$. Strichartz spaces
have become popular in the context of global scattering and asymptotic stability of solitary waves.
Nakanishi \cite{Nak} applied these spaces to the proof of global well-posedness for small initial data
in the nonlinear Klein--Gordon and Schr\"{o}dinger equations. Here we 
consider the nonlinear Dirac equations and modify arguments 
from a more general work of Pelinovsky \& Stefanov \cite{PS} on
the asymptotic stability of small solitary waves.

Let ${\cal H}$ be the one-dimensional Dirac operator in (\ref{Dirac2}) and
$R_{{\cal H}}(\lambda) = ({\cal H} - \lambda I)^{-1}$ be the resolvent operator, defined as a bounded
operator from $L^2(\R)$ to $L^2(\R)$ for any $\lambda \notin \sigma({\cal H})$, where
$$
\sigma({\cal H}) \equiv (-\infty,-1] \cup [1,\infty).
$$

Using Fourier transform,
$$
f(x) = \frac{1}{2\pi} \int_{\R} \hat{f}(k) e^{i k x} dk, \quad \hat{f}(k) = \int_{\R} f(x) e^{-ikx} dx,
$$
the resolvent operator $R_{{\cal H}}(\lambda)$ can be expressed in the explicit form
\begin{equation}
\label{fourier-transform-Dirac}
(R_{{\cal H}}(\lambda) f)(x) = \frac{1}{2 \pi} \int_{\R} \frac{e^{ikx}}{\lambda^2 - 1 - k^2}
\left[ \begin{array}{cc} -(k+\lambda) & 1 \\ 1 & (k-\lambda) \end{array} \right] f(x) dx, \quad
\lambda \notin \sigma({\cal H}).
\end{equation}

Let $\kappa \in \C$ be a solution of the algebraic equation $\kappa^2 + \lambda^2 = 1$
for $\lambda \notin \sigma({\cal H})$ such that ${\rm Re}(\kappa) > 0$. After computations
of the Fourier integrals, the resolvent operator can be expressed in the Green's function form,
\begin{equation}
\label{integral-equations-Volterra}
(R_{{\cal H}}(\lambda) f)(x)  = \frac{1}{2\kappa} \int_{\R}
\left[ \begin{array}{cc} \lambda + i \kappa {\rm sign}(x-y) & -1 \\
-1  & -\lambda - i \kappa {\rm sign}(x-y)
\end{array} \right] e^{-\kappa |x-y|} f(y) dy.
\end{equation}

The Green's function representation (\ref{integral-equations-Volterra}) shows that the resolvent operator
can be extended to the continuous spectrum $\sigma({\cal H})$ as a bounded operator from
$L^1(\R)$ to $L^{\infty}(\R)$ for any $\lambda \in \sigma({\cal H}) \backslash \{1,-1\}$ excluding the end
points $\pm 1$ of the continuous spectrum,
$$
R_{{\cal H}}^{\pm}(\lambda) := \lim_{\epsilon \downarrow 0} R_{{\cal H}}(\lambda \pm i \epsilon), \quad
\lambda \in (-\infty,-1) \cup (1,\infty).
$$
Using the same representation (\ref{integral-equations-Volterra}), we can see
that for any $\lambda_0 > 1$, there is $C(\lambda_0) > 0$ such that
\begin{eqnarray}
\label{b:20} \sup_{|\lambda| \geq \lambda_0} \|  R_{{\cal H}}^{\pm}(\lambda)
\|_{L^1 \to L^\infty} \leq C(\lambda_0).
\end{eqnarray}

Dispersive decay estimates for the semi-group associated to
the one-dimensional Schr\"{o}dinger operator \cite{Weder} extends to those
$e^{-i t {\cal H}}$. Indeed, both semi-groups have the same dispersion for
small Fourier wave numbers $k$. For large Fourier wave numbers, the semi-group
$e^{-i t {\cal H}}$ behaves similar to that of the wave equation and it is
controlled using the uniform asymptotic behavior (\ref{b:20}). As a result,
there is $C > 0$ such that
\begin{equation}
\label{decay-estimates}
\| e^{-i t {\cal H}} f \|_{L^{\infty}} \leq C t^{-1/2} \| f \|_{L^1}.
\end{equation}
Interpolating with the conservation law $\| e^{-i t {\cal H}} f \|_{L^2} = \| f \|_{L^2}$,
we obtain for any $2 \leq p \leq \infty$,
\begin{equation}
\label{decay-estimates-p}
\| e^{-i t {\cal H}} f \|_{L^{p'}} \leq C t^{1/p-1/2} \| f \|_{L^p}, \quad \frac{1}{p} + \frac{1}{p'} = 1.
\end{equation}

Pointwise dispersive decay estimates (\ref{decay-estimates-p}) allow us to
introduce Strichartz admissible pairs and Strichartz estimates for the nonlinear Dirac equations. 

\begin{definition}
\label{defi:1} We say that a pair $(q,r)$ is Strichartz admissible
for the nonlinear Dirac equations if
$$
q \geq 2, \quad r\geq 2 \quad \mbox{\rm and} \quad \frac{2}{q} + \frac{1}{r} \leq \frac{1}{2}.
$$
In particular, $(q,r)=(4,\infty)$ and $(q,r)=(\infty,2)$ are
end-point Strichartz pairs.
\end{definition}

\begin{lemma}
\label{le:1} Let $(q,r)$ be a Strichartz admissible pair. There are constants $C > 0$ such that
 \begin{eqnarray}
 \label{120}
 & &   \|e^{-i t {\cal H}} f\|_{L^{4}_t L^\infty_x} \leq C \|f\|_{H_x^1},  \\
\label{a50}
 & &   \|e^{-i t {\cal H}} f\|_{L^{\infty}_t H^1_x}\leq
 C  \|f\|_{H^1_x},\\
 \label{a:3}
 & &  \left\|\int_0^t e^{-i(t-\tau) {\cal H}}
 F(\tau, \cdot)d\tau \right\|_{L^{4}_t L^\infty_x \cap L^{\infty}_t H^1_x}\leq C
 \|F\|_{L^{1}_t H^1_x}.
 \end{eqnarray}
\end{lemma}

The following theorem simplifies the main result of Pelinovsky \& Stefanov \cite{PS}
to the global small-norm solutions of the nonlinear Dirac equations with quintic and higher-order
nonlinear terms.

\begin{theorem}
\label{theorem-main} Consider the nonlinear Dirac equations (\ref{Dirac2}) with
$$
{\bf f}(a {\bf u}) = a^{2n+1} {\bf f}({\bf u}), \quad a \in \R,
$$
for a fixed integer $n \geq 2$. Assume that ${\bf u}(0) \in H^1(\R)$ and $\| {\bf u}(0) \|_{H^1}$ 
is sufficiently small. The nonlinear Dirac equations (\ref{Dirac2}) admits
a global solution
$$
{\bf u}(t) \in C(\mathbb{R}_+,H^1(\R)) \cap L^4(\mathbb{R}_+,L^\infty(\R)).
$$
\end{theorem}

\begin{proof}
By Duhamel's principle, we can rewrite the Cauchy problem for the nonlinear Dirac equations (\ref{Dirac2})
in the integral form,
\begin{equation}
\label{integral-eq}
{\bf u}(t) = e^{-i t {\cal H}} {\bf u}(0) + \int_0^t e^{-i (t-s) {\cal H}} {\bf f}({\bf u}(s)) ds.
\end{equation}
By Lemma \ref{le:1}, solutions of the integral equation (\ref{integral-eq}) satisfy the bound,
\begin{equation}
\label{e:2} \| {\bf u} \|_{L^4_t L^\infty_x \cap L^\infty_t H^1_x} \leq
C \| {\bf u}_0 \|_{H^1} + C \| {\bf f}({\bf u}) \|_{L^1_t H^1_x},
\end{equation}
for some $C > 0$. We set up the problem of solving the integral equation (\ref{integral-eq})
as an iteration scheme, where we look for a fixed point in a small ball in normed space
$L^4_t L^\infty_x \cap L^\infty_t H^1_x$. Because ${\bf f}({\bf u})$ is a homogeneous polynomial
of degree $2n+1$, we obtain
\begin{eqnarray*}
\|{\bf f}({\bf u}) \|_{L^1_t H^1_x} \leq C \| (|{\bf u}|+ |\partial_x {\bf u}|)|{\bf u}|^{2n}\|_{L^1_t L^2_x}
 \leq C \|{\bf u}\|_{L^\infty_t H^1_x} \|{\bf u}\|_{L^{2n}_t L^{\infty}_x}^{2n}.
\end{eqnarray*}
By Sobolev embedding and the log convexity of the $L^{2n}(\R)$ norms for any $n \geq 2$, we have
$$
\|{\bf u}\|_{L^{2n}_t L^\infty_x}\leq \|{\bf u}\|_{L^{4}_t
L^\infty_x}^{2/n} \|{\bf u}\|_{L^{\infty}_t L^\infty_x}^{1-2/n}
\leq C \|{\bf u}\|_{L^4_t L^\infty_x \cap L^\infty_t H^1_x}.
$$
As a result, we obtain
$$
\|{\bf f}({\bf u})\|_{L^1_t H^1_x} \leq C \|{\bf u}\|_{L^4_t L^\infty_x \cap L^\infty_t H^1_x}^{2n+1},
$$
and the fixed point argument is closed for small ${\bf u}(0) \in H^1(\R)$.
\end{proof}

\begin{remark}
Because $\| {\bf u}(t) \|_{L^{\infty}}$ is a continuous function of $t \in \R_+$ 
and $\| {\bf u}(t) \|_{L^{\infty}} \in L^4(\R_+)$, we have
$$\lim_{t \to \infty} \| {\bf u}(t) \|_{L^{\infty}} = 0.$$
Moreover, $\| {\bf u}(t) \|_{L^{\infty}} = {\cal O}(t^{-1/4-\nu})$ as $t \to \infty$ for some $\nu > 0$.
\end{remark}

\section{Scattering around zero for cubic Dirac equations}

Here we consider the nonlinear Dirac equations (\ref{Dirac1}) with the cubic nonlinear terms,
which are generated by the quartic function $W$ in (\ref{potential}). If $W$ is a function of
$|u|^2$ and $|v|^2$, results of Theorem \ref{theorem-global-wellposedness-cme} show that
the local solutions in $H^1$ are globally well-posed. On the other hand, decay to zero of small
initial data is excluded from results of Theorem \ref{theorem-main} because the cubic nonlinear terms
with $n = 1$ can not be treated by the nonlinear analysis in Strichartz spaces.
Additional constraints must be imposed to ensure that small initial data in $H^1$ scatter to zero.

A similar question has been addressed in the context of the nonlinear Klein--Gordon equation,
\begin{equation}
u_{tt} - u_{xx} + u + |u|^{p-1} u = 0,
\end{equation}
Let us consider the semi-group of the linear Klein--Gordon equation,
$$
S(t) := e^{-it \langle \partial_x \rangle}, \quad t > 0,
$$
where $\langle x \rangle = \sqrt{1 + x^2}$. The $L^{\infty}-L^1$ norm of the semi-group
decays like ${\cal O}(t^{-1/2})$ as $t \to \infty$. As a result, the term $\|u \|_{L^{\infty}}^{p-1}
= {\cal O}(t^{-(p-1)/2})$ is absolutely integrable in $t$ for $p > 3$.
The dispersive decay of small initial data for $p > 3$ was proven by Georgiev \& Lecente \cite{Geor}.
In the critical case $p = 3$ of the cubic nonlinear terms,
the decay of small initial data and the scattering to zero was recently obtained
by Hayashi \& Naumkin \cite{HN1,HN2}.

We will show that the results of Hayashi \& Naumkin \cite{HN1,HN2} can be equally applied
to the nonlinear Dirac equations (\ref{Dirac1}). Using the Fourier transform, we rewrite the system as
\begin{equation}
\label{cme-fourier}
\left\{ \begin{array}{cc} i \hat{u}_t  - k \hat{u} + \hat{v} = \hat{f}, \\
i \hat{v}_t + k \hat{v} + \hat{u} = \hat{g}, \end{array} \right.
\end{equation}
Using the projection matrix
$$
\hat{P} = \left[ \begin{array}{cc} 1 & -\sqrt{1+k^2}-k \\ \sqrt{1+k^2} + k & 1 \end{array} \right], \;\;
\hat{P}^{-1} = \frac{1}{2 \sqrt{1 + k^2}}
\left[ \begin{array}{cc} \sqrt{1+k^2}-k & 1 \\ -1 & \sqrt{1+k^2}-k \end{array} \right],
$$
we can write (\ref{cme-fourier}) in the equivalent form
\begin{equation}
\label{cme-fourier-ab}
\left\{ \begin{array}{cc} i \hat{a}_t  - \langle k \rangle \hat{a} = \frac{1}{2 \langle k \rangle}
(-\hat{f} + (\langle k \rangle - k) \hat{g}), \\
i \hat{b}_t + \langle k \rangle \hat{b} =  \frac{1}{2 \langle k \rangle}
(\hat{g} + (\langle k \rangle - k) \hat{f}), \end{array} \right.
\end{equation}
In the physical space, this system takes the form
\begin{equation}
\label{cme-ab}
\left\{ \begin{array}{l} a_t + i \langle i \partial_x \rangle \hat{a} = \frac{1}{2i}
\langle i \partial_x \rangle^{-1} (-\tilde{f}(a,b) + (\langle i \partial_x \rangle + i \partial_x) \tilde{g}(a,b)), \\
b_t - i \langle i \partial_x \rangle b =  \frac{1}{2i}
\partial_x \rangle^{-1} (\tilde{g}(a,b) + (\langle i \partial_x \rangle + i \partial_x) \tilde{f}(a,b)), \end{array} \right.
\end{equation}
where
\begin{eqnarray*}
\tilde{f}(a,b) & = & f(b + (i \partial_x - \langle i \partial_x \rangle) a, a + (\langle i \partial_x \rangle - i \partial_x) b), \\ \tilde{g}(a,b) & = & g(b + (i \partial_x - \langle i \partial_x \rangle) a, a + (\langle i \partial_x \rangle - i \partial_x) b).
\end{eqnarray*}

If $f$ and $g$ are homogeneous cubic polynomials in variables $u$ and $v$, then $\tilde{f}$ and $\tilde{g}$ are
cubic polynomials in variables $a$, $b$, $\partial_x a$, $\partial_x b$, $\langle i \partial_x \rangle a$, and
$\langle i \partial_x \rangle b$. By Remark 1.1 in \cite{HN2}, all these cubic coefficients can be treated in spaces
$H^m_s(\R)$ equipped with the norm,
$$
\| u \|_{H^m_s} := \| \langle x \rangle^s \langle i \partial_x \rangle^m u \|_{L^2}.
$$
The method of Hayashi \& Naumkin \cite{HN1,HN2} gives the following theorem.

\begin{theorem}
Fix small $\epsilon > 0$ and assume that ${\bf u}_0 \in H^4_1(\mathbb{R})$ with $\| {\bf u}_0 \|_{H^4_1} \leq \epsilon$. There exists $\epsilon_0 > 0$ such that for all $\epsilon \in (0,\epsilon_0)$,
the Cauchy problem for the nonlinear Dirac equations (\ref{Dirac2}) with
the quartic function $W$ in (\ref{potential})
has a unique global solution
$$
{\bf u}(t) \in C(\mathbb{R}_+,H^4_1(\mathbb{R})),
$$
satisfying the time decay estimate
$$
\| \langle i \partial_x \rangle {\bf u}(t) \|_{L^{\infty}} \leq C \epsilon (1 + t)^{-1/2}, \quad t \in \mathbb{R}_+.
$$
\label{theorem-HN}
\end{theorem}

\begin{remark}
Small $H^4_1$ norm on the initial data implies small $H^1$ and $L^1$ norms of ${\bf u}_0$.
\end{remark}

We conclude the article by considering the integrable case of the nonlinear Dirac equations (\ref{Dirac1})
with $W = 2 |u|^2 |v|^2$, which is referred to as the massive Thirring model (MTM).
Theorem \ref{theorem-global-wellposedness-cme} implies global existence of solutions of (MTM) in $H^1$.
Theorem \ref{theorem-HN} implies that these solution scatter to zero in the $L^{\infty}$ norm. More
results were obtained for the massive Thirring model recently.

Selberg and Tesfahun \cite{ST} proved local well-posedness of (MTM) in $H^s(\R)$ for $s > 0$
and global well-posedness in $H^s(\R)$ for $s > \frac{1}{2}$.
Machihara {\em et al.} \cite{Machihara} proved for a similar nonlinear Dirac equations with
quadratic nonlinear terms that local well-posedness holds in $H^s(\R)$ for $s > -\frac{1}{2}$
and that the Cauchy problem is ill-posed in $H^{-1/2}(\R)$.
Using ideas from \cite{ST} and \cite{Machihara}, Candy \cite{Candy} proved local and global
well-posedness of (MTM) in $L^2$.

In characteristic coordinates,
$$
\xi = \frac{x-t}{2}, \quad \tau = \frac{x+t}{2},
$$
the massive Thirring model with $W = 2 |u|^2 |v|^2$ is written explicitly by
\begin{equation}
\label{MTM-end}
\left\{ \begin{array}{cc} i u_{\tau} + v = 2 |v|^2 u, \\
- i v_{\xi} + u = 2 |u|^2 v. \end{array} \right.
\end{equation}

Let us introduce the change of variables,
\begin{eqnarray}
\label{sub-1}
u(\xi,\tau) = \frac{1}{2} \; w(\xi,\tau) \; \exp\left( -\frac{i}{2}
\int_{\xi}^{\infty} |w|^2(\xi',\tau) d\xi'\right).
\end{eqnarray}
The second equation of system (\ref{MTM-end}) can be solved with
\begin{eqnarray}
\label{sub-2}
v(\xi,\tau) & = & -\frac{i}{2} \partial_{\xi}^{-1} w(\xi,\tau) \exp\left(
-\frac{i}{2} \int_{\xi}^{\infty} |w|^2(\xi',\tau) d\xi'\right),
\end{eqnarray}
where
$$
\partial_{\xi}^{-1} w(\xi,\tau) := -\int_{\xi}^{\infty} w(\xi',\tau) d\xi'.
$$
If $v(\cdot,\tau) \in H^1(\R)$, then the zero-mass constraint $\int_{\R} w(\xi,\tau) d\xi = 0$ has
to be added.

With the substitutions (\ref{sub-1})--(\ref{sub-2}) to (\ref{MTM-end}), the massive Thirring model
becomes the scalar evolution equation
\begin{equation}
\label{MTM-scalar}
w_{\tau} - \partial_{\xi}^{-1} w + i |\partial_{\xi}^{-1} w|^2 w = 0.
\end{equation}
The scalar equation (\ref{MTM-scalar}) is invariant under the following change of variables,
\begin{equation}
\label{scaling}
w = \delta W(X,T), \quad X = \delta^2 \xi, \quad T = \delta^{-2} \tau, \quad \delta > 0,
\end{equation}
which implies that the massive Thirring model is the $L^2$-critical model with $\| w \|_{L^2} = \| W \|_{L^2}$.
Therefore, it is natural to expect that the dispersive decay to zero can occur already for 
a smooth initial data with a small $L^2$-norm. To deal with this question, 
we shall review the inverse scattering transform method
for the massive Thirring model.

The scalar equation in characteristic coordinates (\ref{MTM-scalar})
appears as a solvability condition  \cite{KN,KPR} of the spectral problem
\begin{equation}
\label{Lax-1}
\partial_{\xi} \left[ \begin{array}{c} \psi_1 \\ \psi_2 \end{array} \right] =
\left[ \begin{array}{cc} -i \lambda^2 & \lambda w \\ - \lambda \bar{w} &  i \lambda^2 \end{array} \right]
\left[ \begin{array}{c} \psi_1 \\ \psi_2 \end{array} \right]
\end{equation}
and the linear time-evolution problem
\begin{equation}
\label{Lax-2}
\partial_{\tau} \left[ \begin{array}{c} \psi_1 \\ \psi_2 \end{array} \right] = i
\left[ \begin{array}{cc} \eta^2 - \frac{1}{2} |\partial_{\xi}^{-1} w|^2 &  -\eta \partial_{\xi}^{-1} w \\
- \eta \partial_{\xi}^{-1} w &  - \eta^2 + \frac{1}{2} |\partial_{\xi}^{-1} w|^2 \end{array} \right]
\left[ \begin{array}{c} \psi_1 \\ \psi_2 \end{array} \right],
\end{equation}
where $\lambda \in \C$ is a $(\xi,\tau)$-independent spectral parameter and
$\eta = \frac{1}{2\lambda}$. Note that the massive Thirring model in the laboratory coordinates
can also be represented as a solvability condition of the Lax system \cite{KL,KM}.

The spectral problem (\ref{Lax-1}) has some symmetries. If $(\psi_1(x;\lambda),\psi_2(x;\lambda))$ is
a solution of (\ref{Lax-1}), then
\begin{equation}
\label{symmetry}
(\psi_1(x;-\lambda),-\psi_2(x;-\lambda)), \quad (\bar{\psi}_2(x;-\bar{\lambda}),\bar{\psi}_1(x;-\bar{\lambda})),
\quad (\bar{\psi}_2(x;\bar{\lambda}),-\bar{\psi}_1(x;\bar{\lambda}))
\end{equation}
are also solutions of (\ref{Lax-1}).

The continuous spectrum of the spectral problem (\ref{Lax-1}) is located for $\lambda \in \R \cup \{ i \R\}$,
whereas isolated eigenvalues are located symmetrically in quartets $(\lambda,-\lambda,\bar{\lambda},-\bar{\lambda})$ in all quadrants of the complex plane for $\lambda$ \cite{KPR}. Isolated eigenvalues are associated
with solitons that occur in the long-time dynamics of the solution $w(\xi,\tau)$ thanks
to the independence of $\lambda$ from $\tau$ and the inverse scattering transform technique. We shall prove
that solitons are absent if $w(\cdot,\tau)$ for a frozen $\tau$ has a small norm in
$L^2(\R) \cap L^{\infty}(\R)$ (or $H^1(\R)$). For clarity of presentation, we do not write
$\tau$ in the arguments of $w$ and $\psi_{1,2}$.

\begin{lemma}
Fix small $\epsilon > 0$ and assume that $w \in L^2(\mathbb{R})$ with $\| w \|_{L^2} \leq \epsilon$.
There is $C > 0$ such that the spectral problem (\ref{Lax-1}) admits no solutions in $L^2(\R)$ for any
$\lambda \in \C$ with $\arg(\lambda) \in \left( C \epsilon^2, \frac{\pi}{2} - C \epsilon^2 \right)$.
If in addition, $w \in L^1(\R) \cap L^{\infty}(\R)$ and $\partial_{\xi} w \in L^1(\R)$ with
\begin{equation}
\label{condition-eigenvalue}
\| w \|_{L^1} ( \| w \|_{L^{\infty}} + \| \partial_{\xi} w \|_{L^1}) \leq \epsilon,
\end{equation}
then the spectral problem (\ref{Lax-1}) admits no solutions in $L^2(\R)$ for any
$\lambda \in \C$.
\label{lemma-Lax}
\end{lemma}

\begin{proof}
Let us choose $\lambda \in \C$ in the first quadrant of the complex plane.
Because of the symmetry (\ref{symmetry}) of eigenvectors, the results
are valid in all four quadrants of the complex plane. If $\lambda$ is
in the first quadrant of $\C$, then ${\rm Im}(\lambda^2) > 0$ and
we can parameterize $\lambda$ by $\lambda = |\lambda| e^{i \theta}$
with $\theta \in \left(0,\frac{\pi}{2}\right)$.

If ${\rm Im}(\lambda^2) > 0$, then $e^{-i \lambda^2 \xi}$ decays to zero as $\xi \to -\infty$, and
we can introduce $\mbox{\boldmath $\psi$}(\xi) = e^{-i \lambda^2 \xi} \mbox{\boldmath $\varphi$}(\xi)$
with boundary conditions $\lim_{\xi \to -\infty} \mbox{\boldmath $\varphi$}(\xi) = (1,0)^T$
for eigenvectors of the spectral problem (\ref{Lax-1}).

Integrating the system from $-\infty$ to $\xi$ under the boundary conditions for $\mbox{\boldmath $\varphi$}(\xi)$, we obtain the integral equations,
\begin{eqnarray}
\label{integral-system}
\varphi_1(\xi) = 1 + \lambda \int_{-\infty}^{\xi} w(\xi')  \varphi_2(\xi') d\xi', \quad
\varphi_2(\xi) = - \lambda \int_{-\infty}^{\xi} e^{2 i \lambda^2 (\xi - \xi')} \bar{w}(\xi')
\varphi_1(\xi') d\xi'.
\end{eqnarray}

Using the exact integral
$$
I(\lambda) := |\lambda^2| \int_0^{\infty} e^{-2 {\rm Im}(\lambda^2) x} dx
= \frac{|\lambda|^2}{2 {\rm Im}(\lambda^2)} = \frac{1}{2 \sin(2\theta)},
$$
the Schwarz inequality, and Young's inequality for convolution integrals, we obtain
$$
\| \varphi_1 - 1 \|_{L^{\infty}} \leq \| w \|_{L^2} \| \lambda \varphi_2 \|_{L^2}, \quad
\| \lambda \varphi_2 \|_{L^2} \leq I(\lambda) \| w \|_{L^2}
\| \varphi_1 \|_{L^{\infty}}.
$$
Closing the inequalities and using fixed-point arguments, we can see that if
$$
I(\lambda) \| w \|_{L^2}^2 < 1,
$$
then there is a unique solution of system (\ref{integral-system})
for $\varphi_1 \in L^{\infty}(\R)$ and $\lambda \varphi_2 \in L^2(\R)$ such that
$\| \varphi_1 - 1 \|_{L^{\infty}} < 1$. Therefore, $\varphi_1(\xi) \nrightarrow 0$
as $\xi \to +\infty$, and so $\psi_1(\xi)$ grows exponentially as $\xi \to +\infty$.
Eigenvectors in $L^2$ may only exist if $I(\lambda) \| w \|_{L^2}^2 \geq 1$, that is,
if either $\theta \in (0,C \epsilon^2)$ or $\theta \in \left( \frac{\pi}{2}-C \epsilon^2, \frac{\pi}{2}\right)$
for some $C > 0$.

To eliminate eigenvectors everywhere in the first quadrant of $\C$, we add now the condition (\ref{condition-eigenvalue}). Integrating the second equation of system (\ref{integral-system}) by parts
and using the first equation, we obtain
\begin{eqnarray*}
\lambda \varphi_2(\xi) & = & - \lambda^2 \int_{-\infty}^{\xi} e^{2 i \lambda^2 (\xi - \xi')} \bar{w}(\xi')
\varphi_1(\xi') d\xi' \\ & = & \frac{1}{2i} \bar{w}(\xi) \varphi_1(\xi) - \frac{1}{2i}
\int_{-\infty}^{\xi} e^{2 i \lambda^2 (\xi - \xi')} \left( \partial_{\xi'} \bar{w}(\xi')
\varphi_1(\xi') + \bar{w}(\xi') \partial_{\xi'}
\varphi_1(\xi')\right) d\xi' \\
& = & \frac{1}{2i} \bar{w}(\xi) \varphi_1(\xi) - \frac{1}{2i}
\int_{-\infty}^{\xi} e^{2 i \lambda^2 (\xi - \xi')} \partial_{\xi'} \bar{w}(\xi')
\varphi_1(\xi') d\xi' - \frac{1}{2i}
\int_{-\infty}^{\xi} e^{2 i \lambda^2 (\xi - \xi')} |w(\xi')|^2 \lambda
\varphi_2(\xi') d\xi'.
\end{eqnarray*}
Using H\"{o}lder's inequality and Young's inequality for convolution integrals, we obtain
\begin{eqnarray*}
\| \varphi_1 - 1 \|_{L^{\infty}} & \leq & \| w \|_{L^1} \| \lambda \varphi_2 \|_{L^{\infty}}, \\
\| \lambda \varphi_2 \|_{L^{\infty}} & \leq & \frac{1}{2} \left( \| w \|_{L^{\infty}} + \| \partial_{\xi} w \|_{L^1}
\right) \| \varphi_1 \|_{L^{\infty}} + \frac{1}{2} \| w \|^2_{L^2}
\| \lambda \varphi_2 \|_{L^{\infty}}.
\end{eqnarray*}
If $\| w \|^2_{L^2} < 2$, then there is $C > 0$ such that
\begin{eqnarray*}
\|\varphi_1 - 1 \|_{L^{\infty}} & \leq & C \| w \|_{L^1} ( \| w \|_{L^{\infty}} + \| \partial_{\xi} w \|_{L^1}) \| \varphi_1 \|_{L^{\infty}}, \\ \| \lambda \varphi_2 \|_{L^{\infty}} & \leq & C ( \| w \|_{L^{\infty}} + \| \partial_{\xi} w \|_{L^1}) \| \varphi_1 \|_{L^{\infty}}.
\end{eqnarray*}
Under the condition (\ref{condition-eigenvalue}), there is a unique solution of system (\ref{integral-system})
for $\varphi_1 \in L^{\infty}(\R)$ and $\lambda \varphi_2 \in L^{\infty}(\R)$ such that
$\| \varphi_1 - 1 \|_{L^{\infty}} < 1$. Repeating the arguments above, we conclude the proof that no eigenvector in $L^2$ exists for any $\lambda \in \C$ under the condition (\ref{condition-eigenvalue}).
\end{proof}

\begin{remark}
Using the scaling transformation (\ref{scaling}), we can see that both $\| w \|_{L^2}$ and
$\| w \|_{L^1} ( \| w \|_{L^{\infty}} + \| \partial_{\xi} w \|_{L^1})$ are invariant with respect
to parameter $\delta$. Soliton solutions of the massive Thirring model are supported by
particular values for these quantities. If $\| w \|_{L^2}$ and
$\| w \|_{L^1} ( \| w \|_{L^{\infty}} + \| \partial_{\xi} w \|_{L^1})$ are below these 
particular values, no solitons can occur in the
long-time evolution of the massive Thirring model.
\end{remark}

Further analysis of the inverse scattering transform using the time-evolution problem (\ref{Lax-2})
may give an analogue of Theorem \ref{theorem-HN} for the massive Thirring model,
perhaps, with relaxed assumptions on the initial data ${\bf u}_0$. Another interesting open problem
is to explore global existence of the massive Thirring model in $L^2$ \cite{Candy}
and obtain $L^2$-orbital stability of MTM solitons. A similar task was recently achieved
by Mizumachi \& Pelinovsky \cite{MP} in the context of the nonlinear Schr\"{o}dinger equation.
The massive Thirring model is more interesting for orbital stability analysis of solitary waves.
Because it is associated with the sign-indefinite Hamiltonian function (\ref{quantity-H}),
no orbital stability in $H^1$ can be extracted from the standard energy analysis.
These open problems will likely to attract interests of researchers in near future.

\end{document}